\documentclass[submission,copyright,creativecommons]{eptcs}
\providecommand{\event}{Graph Computation Models (GCM) 2020} 
\usepackage{breakurl}             
\usepackage{underscore}           
\usepackage{amssymb}
\usepackage{pifont}
\usepackage{graphicx}
\usepackage{acronym}
\usepackage[utf8]{inputenc}
\usepackage{todonotes}
\usepackage{floatflt} 
\usepackage{color}
\usepackage{url}
\usepackage{float}

\usepackage{chngcntr}

\newacro{MDE}{Model-Driven Engineering}
\newacro{TGG}{Triple Graph Grammar}
\newacro{bx}{bidirectional transformation}
\newacro{GPL}{general purpose language}
\newacro{DSL}{Domain-Specific Language}
\newacro{fUML}{Foundational UML}
\newacro{DEVS}{discrete event system specifications}
\newacro{ILP}{Integer Linear Programming}
\newacro{UI}{user interface}
\newacro{IDE}{integrated development environment}
\newacro{XMI}{XML metadata interchange}
\newacro{AADL}{Architecture Analysis and Design Language}

\title{VICToRy: Visual Interactive Consistency Management in Tolerant Rule-based Systems}
\author{Nils Weidmann
	\institute{Paderborn University \\ Paderborn, Germany}
		\email{nils.weidmann@upb.de} 
		\and
	Anthony Anjorin
	\institute{IAV GmbH Ingenieurgesellschaft Auto und Verkehr \\ Berlin, Germany}
		\email{anthony.anjorin@iav.de}
		\and
	James Cheney
	\institute{University of Edinburgh \\ Edinburgh, United Kingdom}
		\email{jcheney@inf.ed.ac.uk}
	}

\def\titlerunning{VICToRy: Visual Interactive Consistency Management in Tolerant Rule-based Systems}
\def\authorrunning{N. Weidmann, A. Anjorin \& J. Cheney}

\begin{document}

\maketitle              

\begin{abstract}
	%
	In the field of Model-Driven Engineering, there exist numerous tools that support various consistency management operations including model transformation, synchronisation and consistency checking.
	%
	%
	The supported operations, however, typically run completely in the background with only input and output made visible to the user.
	We argue that this often reduces both understandability and controllability.
	%
	%
	As a step towards improving this situation, we present VICToRy, a debugger for model generation and transformation based on Triple Graph Grammars, a well-known rule-based approach to bidirectional transformation.
	In addition to a fine-grained, step-by-step, interactive visualisation, VICToRy enables the user to actively explore and choose between multiple valid rule applications thus improving control and understanding.
	
\end{abstract}


%
%
%
\section{Introduction and Motivation}
\label{sec:intro}

As an approach to model management, \ac{bx} enables designers to derive multiple consistency management operations from a single specification.
Concepts of \ac{bx} for various research fields exist, including \ac{MDE}, databases, and programming languages.
Among other formal approaches, \acp{TGG} are a rule-based approach to \ac{bx}, in which the consistency relation between two models (interchangeably denoted as source and target) is expressed by a third correspondence model.
A convenient feature of \acp{TGG} is that declarative rules are used as a common basis from which various operations such as unidirectional transformation, model synchronisation and consistency checking are derived.
This can, however, be confusing for designers who are not aware of or used to the underlying derivation process.
While some \ac{TGG} tools provide basic debugging functionality for the transformation process (cf. Sect.~\ref{sec:related}), none of them enable the user to track let alone influence the choice of rule applications.
In most cases, only the input and output models are visible to the user, whereas the transformation process runs in the background.
Stevens~\cite{Stevens2014} argues for more transparent and fault-tolerant model transformation approaches, such that the user should be involved in controversial decisions, i.e. decisions that cannot be made only based on the consistency relation specification.
Even for uncontroversial transformations, we have observed that novice users are unable to fully understand how \ac{TGG} tools - viewed as black-boxes - determine a specific result.

In this paper, we propose the \emph{VICToRy} debugger\footnote{\url{https://github.com/eMoflon/emoflon-victory}} as an add-on component for supporting an interactive step-by-step visualisation of model transformations based on \acp{TGG}. 
It presents possible operational rules including their concrete application contexts to the user, as well as a history of the involved models as they evolve during the transformation process.
Additionally, the user can inspect and choose a valid rule application at each time step, or decide to resume the automated process in the background.
Up to now, the operations model generation and forward/backward transformation are supported, while model synchronisation and consistency checking are planned to be implemented soon.
VICToRy is currently integrated into the eMoflon tool suite\footnote{\url{https://emoflon.org/}} but can be potentially connected to other Java-based \ac{TGG} and even general graph transformation tools via the defined interfaces.
This means that existing and future tools can be enriched with debugging facilities to increase user involvement and understanding in the transformation process, and thereby contributing to fault-tolerant model management.
Previous work~\cite{Cheney2012,Anjorin2019} identified possible connections between \ac{bx} and \textit{provenance}, to which our approach makes a further practical contribution.
By providing access to why-provenance (i.e. why is a given pair of models consistent?) and how-provenance (i.e. how it can be proven that two models are consistent?) that \ac{bx} tools produce and maintain, we strive to improve the understandability of such tools.
For \ac{TGG} tools, it was identified that why-provenance is provided by the resulting correspondence model, and how-provenance by the underlying sequence of rule applications.
While this knowledge helps to understand \ac{TGG} tools conceptually, it is also important to provide interactive access to these data structures.  
VICToRy visualises both the correspondence model and the sequence of rule applications, allowing users to actively inspect and shape the emerging provenance structures.

The remainder of this paper is structured as follows:
An overview of supported features is presented in Sect.~\ref{sec:features} using a running example.
The software architecture of VICToRy is sketched in Sect.~\ref{sec:architecture}.
A brief overview of a first case study with VICToRy is provided in Sect.~\ref{sec:evaluation}.
In Sect.~\ref{sec:related} related work is discussed, while Sect.~\ref{sec:conclusion} concludes the paper.
\section{An Overview of Supported Features}
\label{sec:features}

This section provides an overview of features of VICToRy that can help novice users explore an unknown \ac{TGG}.
To identify these features, we conducted an explorative study with students who did not have any \ac{MDE}-related experience at that point (cf. Sect.~\ref{sec:evaluation} for details).

\subsection{Running Example}

\begin{floatingfigure}[r]{0.45\columnwidth}
	\centering
	\includegraphics[width=0.45\columnwidth]{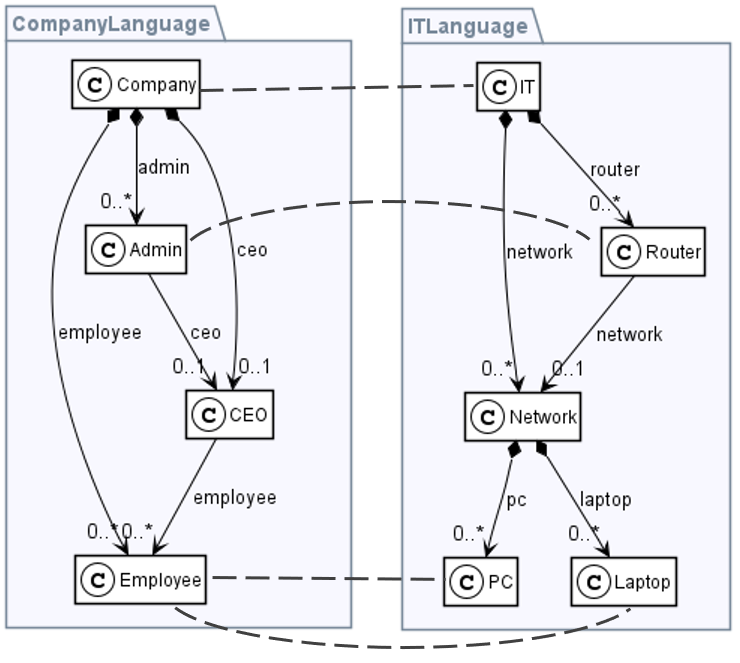}
	\vspace{-5mm}
	\caption{Metamodels: \texttt{CompanyToIT}}
	\label{fig:metamodels}
\end{floatingfigure}

To demonstrate different features of VICToRy, the \ac{bx} example \emph{CompanyToIT} is used,\footnote{\url{http://bx-community.wikidot.com/examples:companytoit}} which maintains consistency between a simplified organisational structure of a company and its corresponding IT infrastructure.
The two metamodels are depicted in Fig.~\ref{fig:metamodels}.
A \texttt{Com\-pany} consists of multiple \texttt{Admin}s, \texttt{Employee}s, and at most one \texttt{CEO}. 
Each \texttt{Admin} reports to the \texttt{CEO} of the \texttt{Com\-pany}, just like the other \texttt{Employees}.
The corresponding \texttt{IT}  involves a set of \texttt{Router}s and \texttt{Net\-work}s; a \texttt{Router} is always assigned to a particular \texttt{Network}.
The \texttt{Network} itself consists of a set of \texttt{PC}s and a set of \texttt{Laptop}s.
A \texttt{Company} and an \texttt{IT} correspond to each other; this is depicted by a dashed line connecting the two nodes in the metamodel.
The same holds for an \texttt{Admin} who is responsible for a \texttt{Router} within the IT infrastructure.
Finally, an \texttt{Employee} can either work with a \texttt{PC} or a \texttt{Laptop}, leading to two correspondence links for the \texttt{Employee} in the metamodel.\\

\subsection{Configurable Visualisation of Rules and Rule Applications}
\label{sec:visualisation-section}

\begin{figure}[H]
	\centering
	\includegraphics[width=\columnwidth]{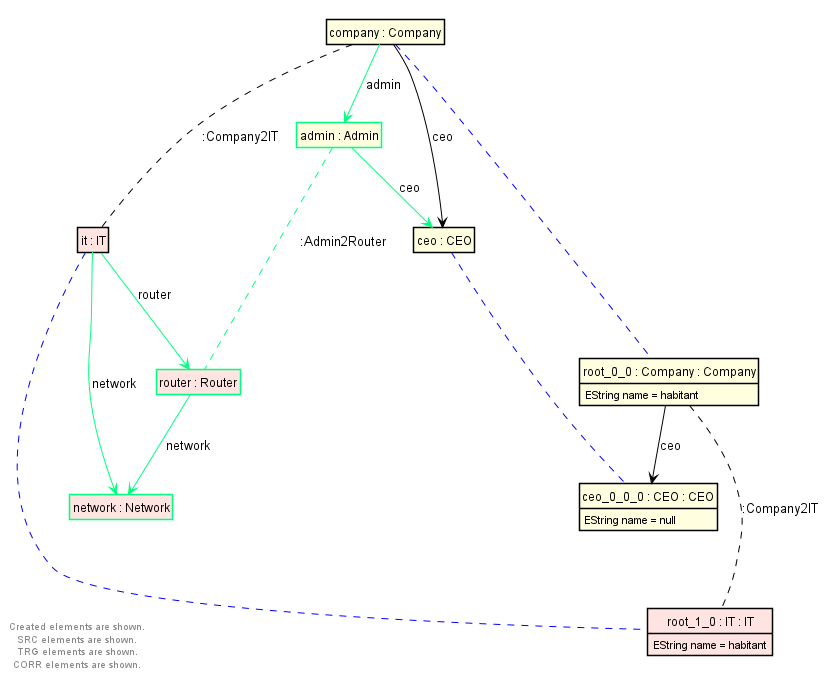}
	\vspace{-10mm}
	\caption{Visualising rules and matches}
	\label{fig:visualisation-section}
\end{figure}

To understand the effects of a rule application on a concrete model, it is essential to visualise both the rule and the resulting model changes at runtime.
The example rule \texttt{AdminTo\-Router}, which creates an \texttt{Admin} in the \texttt{Company} model, and relates them to a created \texttt{Router} in a created \texttt{Network} in the \texttt{IT} model, is depicted to the left of Fig.~\ref{fig:visualisation-section}. 
Created (context) elements of the rule have a green (black) outline.

A specific match of this rule is depicted to the right of Fig.~\ref{fig:visualisation-section}.
Variables in the rule and corresponding matched elements in the model are connected with dashed purple lines. 
In both the rule and its match, the background colour of source model elements is peach, while target model elements have a rose background.
Correspondences are represented as dashed black lines.
The visualisation of rules and rule applications is based on PlantUML\footnote{\url{https://plantuml.com/}} and is generated automatically on rule and match selection (cf. Sect.~\ref{sec:rules-section}).
Editing rules is only possible in the underlying \ac{TGG} tool, meaning that rules cannot be adapted at runtime.
As source, target and correspondence models are depicted as a connected triple, the visualisation represents why-provenance for \acp{TGG}.
To cope with a wide range of size and complexity of \ac{TGG} rules, models' sizes, and the user's proficiency, it is crucial to be able to configure the visualisation:

\begin{figure}[htb]
\begin{minipage}{0.6\columnwidth}
	\begin{description}
		\item[Choice of displayed elements:]
		For each domain (source, target, correspondence), the user can hide the respective elements. 
		For rules, it is also possible to display only  context elements and thus focus on the structure required for a match of that rule in the model instance.
		\item[Abbreviation of labels:]
		For nodes, edges and correspondences, it is possible to display the labels completely, in an abbreviated form containing the first and last three letters, or not at all.
		\item[Neighbourhood of matches:]
		As models of realistic size can become much too large to be completely displayed within the debugger, only the match of a selected rule application and a configurable neighbourhood of this match is displayed.
		The distance of a node to the match is defined as the shortest path from this node to any node contained in the match; nodes in the match itself are assigned a distance of 0.
		The \textit{k}-neighbourhood of a match contains all nodes that are at most $k \in [0;3]$ away.
	\end{description}
\end{minipage}
\hfill
\begin{minipage}{0.35\columnwidth}
		\centering
		\includegraphics[width=0.9\columnwidth]{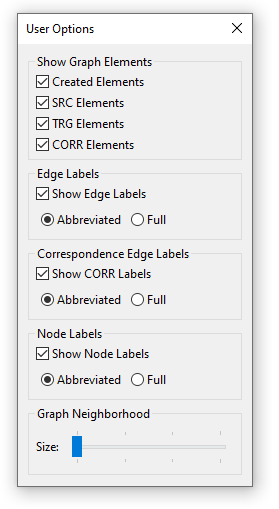}
		\caption{Configuration}
		\label{fig:options}
\end{minipage}
\vspace{-5mm}
\end{figure}

\subsection{Explorable and Interactive Overview of Applied Rules}
\label{sec:rules-section}

\begin{floatingfigure}[r]{0.45\columnwidth}
	\centering
	\includegraphics[width=0.45\columnwidth]{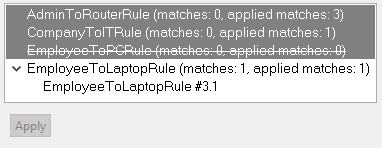}
	\caption{Available rules and matches}
	\label{fig:rules-section}
\end{floatingfigure}

The debugger's \ac{UI} provides an over\-view of all rules of the \ac{TGG}. 
Our example \ac{TGG} CompanyToIT consists of only four rules, which are depicted as a list.
For each rule in the list, the number of available matches in the current model and the number of applied matches are displayed together with the name of the rule.
Rules with a dark grey background are not applicable in the current state of the model, whereas rules with a white background have at least one applicable match.
This provides a quick overview and is useful for \acp{TGG} with a large number of rules.
Furthermore, rules that have never been applicable are crossed out, providing a quick visual indication of rules that might be problematic. 
All matches of a rule can be viewed as sub-entries by expanding the corresponding rule entry in the list.
The currently selected rule and match are visualised in the UI (cf. Sect.~\ref{sec:visualisation-section}).
To apply a rule, the user can either double-click on a particular match, select the match and press the apply button, or simply double-click the rule to apply a random match.
This action is delegated to the connected \ac{TGG} tool, which must handle the actual rule application (cf. Sect.~\ref{sec:architecture}).
As soon as VICToRy receives a response, the \ac{UI} is updated to reflect the new state of the model and available matches.\\


\begin{floatingfigure}[r]{0.45\columnwidth}
	\centering
	\includegraphics[width=0.45\columnwidth]{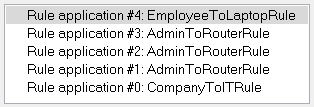}
	\caption{The protocol section}
	\label{fig:protocol-section}
\end{floatingfigure}

The VICToRy debugger provides traceability information by keeping track of all previous rule applications.
This sequence of rule applications is referred to as the (transformation) protocol and represents how-provenance~\cite{Anjorin2019}.
For each protocol entry, the name of the rule as well as a unique ID for the rule application is displayed.
If a protocol entry is selected, the state of the model as created by all rule applications up to and including the selected one is displayed with a configurable neighbourhood; elements created by the selected rule application are highlighted green.
It is also possible to select multiple entries: the respective rule applications are then combined into a single step and visualised accordingly. \\
\section{Architecture}
\label{sec:architecture}

VICToRy can be connected to different Java-based \ac{TGG} tools by implementing a simple interface for transferring data between the debugger and the respective tool.
An overview of this interface is depicted in Fig.~\ref{fig:interface}.
The central component of this interface is the \texttt{Graph} class, which consists of \texttt{Node}s and \texttt{Edge}s.
There exists a mapping from each \texttt{Edge} to a source and a target \texttt{Node}, reflecting the categorical approach to graph transformation.
Both \texttt{Rule}s and \texttt{Matches} are represented as \texttt{Graph}s consisting of rule and model elements, respectively.

\begin{figure}[htb]
	\centering
	\includegraphics[width=0.8\columnwidth]{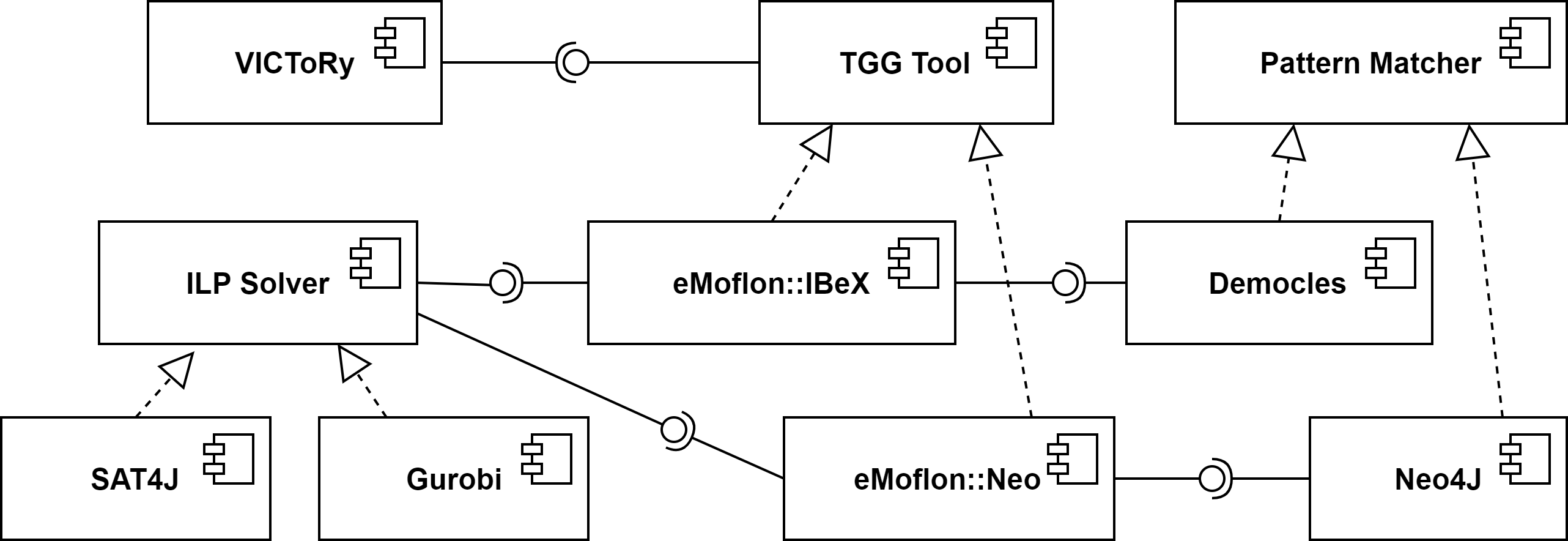}
	\caption{Integrating VICToRy into the eMoflon tool suite}
	\label{fig:component-diagram}
\end{figure}

\begin{floatingfigure}[r]{0.55\columnwidth}
	\centering
	\includegraphics[width=0.55\columnwidth]{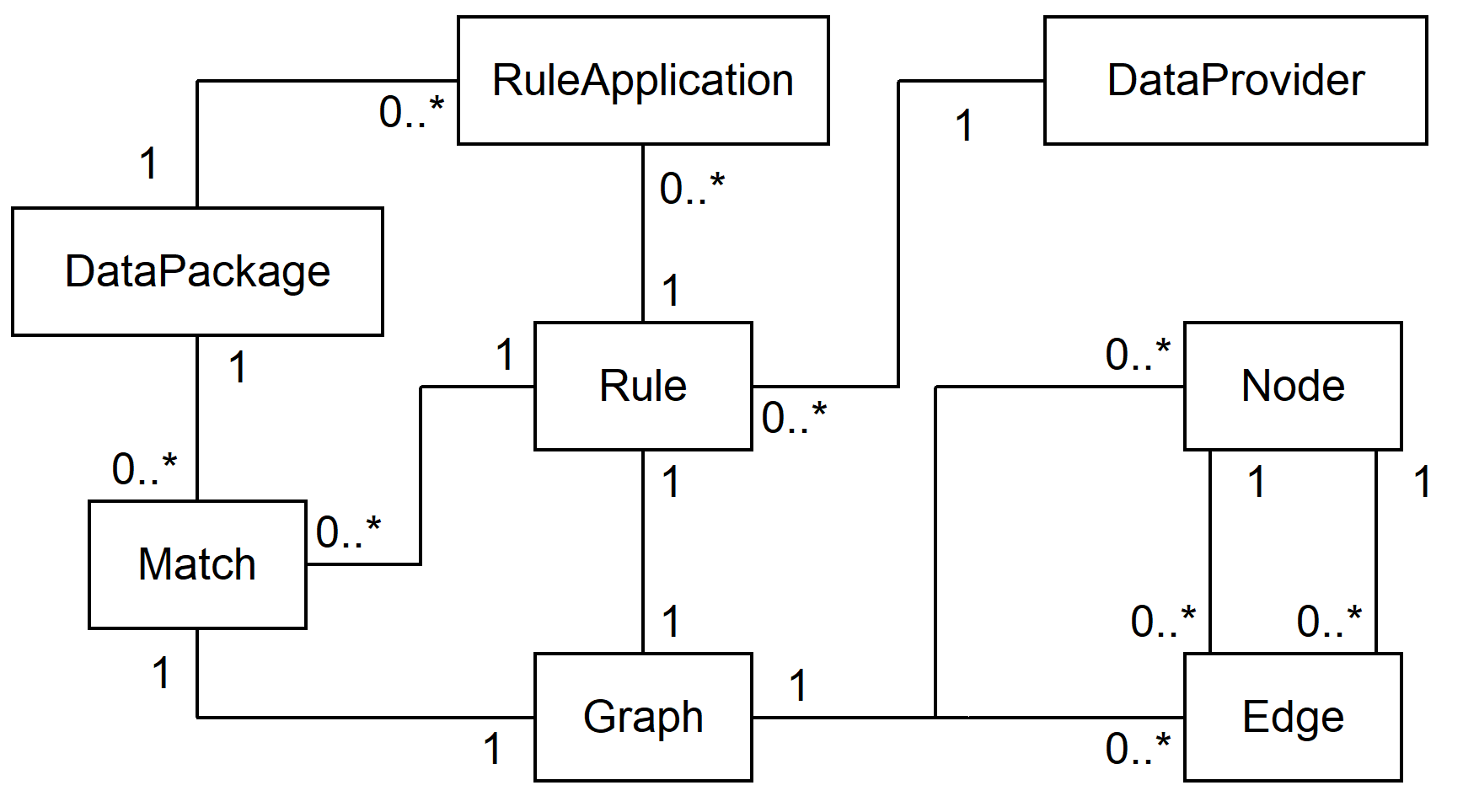}
	\vspace{-5mm}
	\caption{Data exchange with VICToRy}
	\label{fig:interface}
\end{floatingfigure}
Multiple \texttt{Matches} can be determined for the same \texttt{Rule}.
Furthermore, a \texttt{Rule\-Applica\-tion} object is created when a \texttt{Rule} is applied for a concrete \texttt{Match}.
All this information is stored in a \texttt{Data\-Package} transferred between the attached \ac{TGG} tool and VICToRy after each transformation step.
Further information is stored in the \texttt{Data\-Provider},  including all current models and the current choice between debugging and background modi.
As all information necessary to derive the current model state from the initial models and the given \ac{TGG} rules are provided, the interface represents a data exchange metamodel for why- and how-provenance between VICToRy and the attached \ac{TGG} tool.
The component diagram in Fig.~\ref{fig:component-diagram} describes how the debugger has been embedded into the eMoflon tool suite.
VICToRy is designed for - but not limited to - being connected to a \ac{TGG} component of eMoflon.
Currently, both eMoflon::IBeX~\cite{Weidmann2019} and eMof\-lon::Neo\footnote{\url{https://github.com/eMoflon/emoflon-neo}} implement the interface to the debugger.
eMoflon::IBeX uses the incremental pattern matcher Democles~\cite{Varro2012} for matching patterns to graph instances which are loaded from \ac{XMI} files into main memory, whereas eMoflon::Neo uses the external graph database Neo4J to collect matches and store models.
Depending on the operation, both tools are able to construct a superset of possible rule applications, encode the graph problem into an optimisation problem and let an external \ac{ILP} solver determine the final set of rule applications.
Interfaces to the solvers Gurobi\footnote{\url{https://www.gurobi.com/}} and SAT4J\footnote{\url{http://www.sat4j.org/}} are currently implemented, while other solvers can be connected to the tool suite as well.

While performing a consistency management task with VICToRy, the tool switches between two modi as depicted in Fig.~\ref{fig:state-chart}.
In the \textit{background mode}, possible matches for rules are collected and one of them is chosen to be applied.

\begin{floatingfigure}[r]{0.45\columnwidth}
	\centering
	\includegraphics[width=0.45\columnwidth]{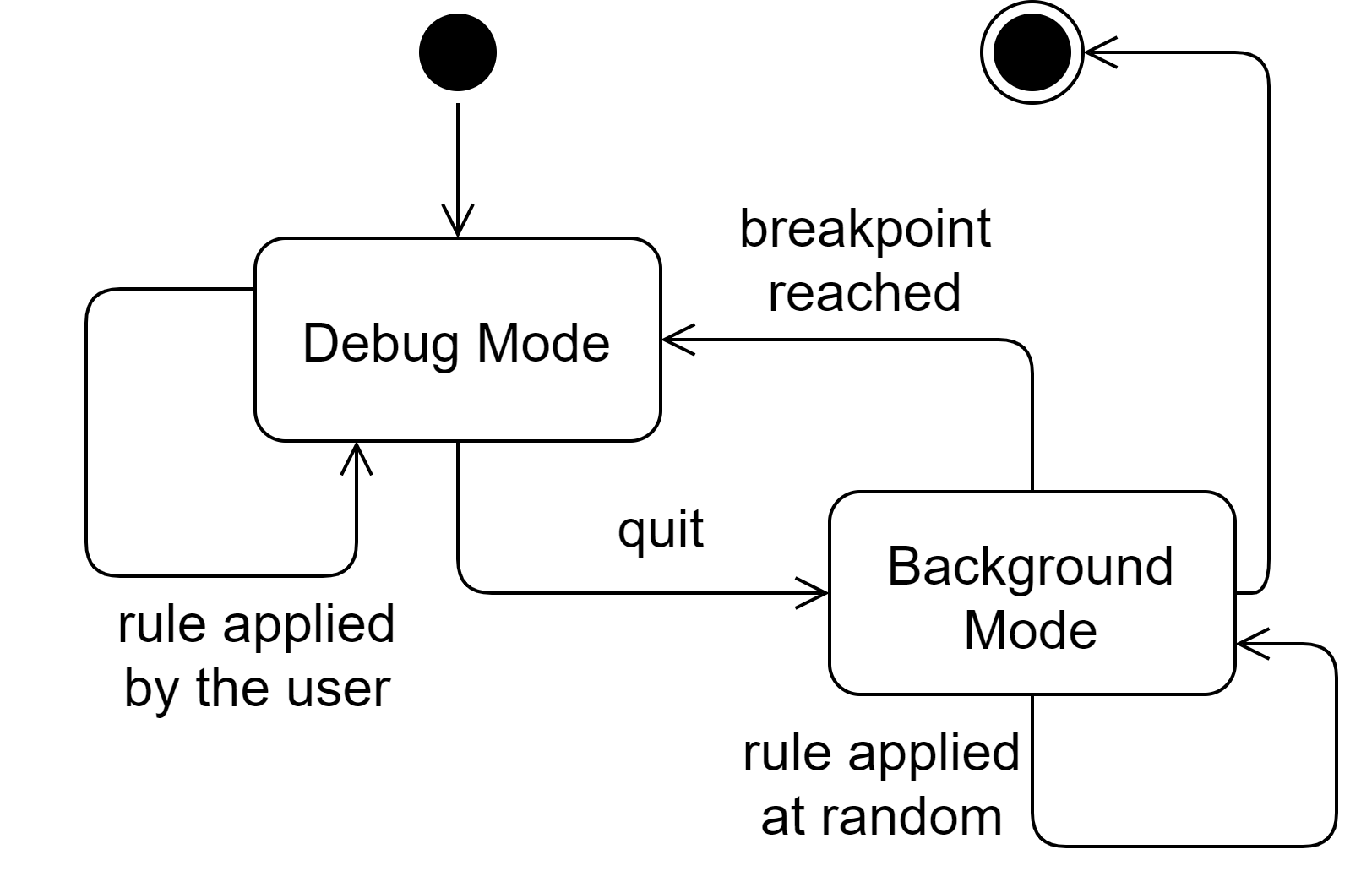}
	\vspace{-5mm}
	\caption{State chart for debugging modi}
	\label{fig:state-chart}
\end{floatingfigure}

In case of multiple options, rule applications are chosen at random (or according to a configurable component) without user interaction, which is the usual workflow for model transformation tools.
This procedure is repeated until no further matches can be found (leading to the termination of the process) or until a breakpoint is reached.
In the latter case, the tool switches to the \textit{debug mode}, where the VICToRy \ac{UI} is visible and each rule application requires a user interaction.
In contrast to the background mode, the user can choose between multiple options which are shown on the \ac{UI}.
To return to the background mode, the user resumes the automated choice of rule application by quitting the debugger.
This behaviour is similar to debugging concepts in contemporary \acp{IDE}, but without the possibility of stepping \emph{into} a rule application.

\section{Case Study}
\label{sec:evaluation}

In order to qualitatively evaluate the usefulness of VICToRy, we conducted a case study adapted from Blouin et al.~\cite{Blouin2014}.
The \ac{AADL} is a standard language in the aerospace domain, for which a textual editor (OSATE) and a graphical editor (Adele) exist.
Blouin et al.~\cite{Blouin2014} discuss the challenge of implementing a bx to synchronise models edited using the different editors.
Our evaluation aims at answering two research questions related to motivational aspects for \ac{MDE} debuggers:

\begin{enumerate}
	\item Does VICToRy help to explore and understand a \ac{TGG} of realistic size?
	\item Does VICToRy help to identify bugs in rules or in input models? 
\end{enumerate} 

\subsection{The Adosate \ac{TGG}}
Blouin et al. implemented the original Adosate \ac{TGG} that maintains consistency between Adele and OSATE models.
The \ac{TGG} consists of 60 rules specified with the model transformation tool MoTE~\cite{Hildebrandt2011}, in which an extended \ac{TGG} formalism is used that allows the designer to connect more than one element per model with a single correspondence.
For the implementation with eMoflon, we created multiple correspondences for each pair of involved source and target nodes.
While the eMoflon \ac{TGG} is much larger than with MoTE, its rule refinement feature~\cite{Anjorin2014} was used to keep the rules manageable.
The feature allowed us to define rules involving abstract node types (abstract rules) that can be refined by concrete types and enriched with additional elements in so-called concrete rules.
These changes resulted in a semantically equivalent \ac{TGG} (i.e., a \ac{TGG} that generates the same language) with 49 abstract rules and 91 concrete rules, of which only the concrete rules are considered at runtime.
An overview comparing the (concrete) rules required for the implementations with MoTE and eMoflon is provided in Table~\ref{tab:rule-groups}.

	\begin{table}[htb]
		\centering
		\begin{tabular}{|c|c|c|c|c|c|}
			\hline
			& \multicolumn{2}{c|}{Number of Rules} & & \multicolumn{2}{c|}{Number of Rules}\\
			AADL Construct & MoTE & eMoflon & AADL Construct & MoTE & eMoflon\\
			\hline
			Package (axiom) & 1 & 2 & Component Type Features & 10 & 19 \\
			Subcomponents & 11 & 12 & Feature Group Types & 4 & 4 \\
			Component Types & 2  & 13 & Feature Group Type Features & 10 & 9 \\
			Connections & 20 & 21 & Component Implementation & 2 & 9 \\
			\hline
			& & & \textbf{Total} & \textbf{60} & \textbf{91} \\
			\hline
		\end{tabular}
		\caption{Size of rule groups: MoTE and eMoflon}
		\label{tab:rule-groups}
	\end{table}

Compared to most \ac{bx} benchmark examples\footnote{\url{http://bx-community.wikidot.com/examples:home}} the number of rules is relatively large, while the average rule size is comparable to existing benchmarks.
The mean of the number of nodes involved in abstract and concrete rules is 6.51 (2.69 created nodes, 3.81 context nodes), and the mean number of edges is 3.31 (2.46 created edges, 0.84 context edges) per rule.

\subsection{Experiment and Results}
In an experiment conducted at Paderborn University, Germany, with 15 computer science graduate students without substantial prior experience with MDE, we attempted to assess if and how VICToRy helps novice users understand a provided, non-trivial TGG.  
The first task was to identify relations between rules and model elements (which model elements are created/required by which rules?), as well as relations between the rules themselves (which rules depend on other rules?).
The students were provided with the Adosate TGG and VICToRy, and asked to work independently on developing an understanding for the TGG.
The first and second authors provided helpful material (tutorials, handbooks, and relevant papers), supervision (answering any basic questions), and held a feedback meeting after two weeks to check the students' understanding for the TGG and ask if VICToRy was helpful and in what ways it was used.  
In general, all students stated that the debugger indeed helped to get an overview of the entire \ac{TGG}.
It was especially helpful to identify which rules are applicable for an empty model (axioms) without clicking through all of them, and to determine which rules provide context for other rules (rule dependencies).
Most students started exploring the \ac{TGG} rules by generating consistent model triples and inspecting the resulting transformation protocol.
In a second step, they then attempted to transform smaller instances in forward and backward directions.
As all models can be saved to disk at any point of time, it was easily possible to try out different alternatives starting from a common state of all models.

In a second task, the students were provided with the Adosate TGG, and with a test suite consisting of input models and expected output models for both forward and backward directions.
One of the students was then asked to either make a change to a TGG rule, or to the supplied input models, resulting in both cases in a mismatch between \ac{TGG} and test suite.
The other students were then asked to determine and explain this mismatch.
This task was carried out in a slightly more controlled manner, restricting the allotted time to a few hours, and asking the students not to perform a diff between the original and changed rules.
The feedback from the students for this task (based again on a feedback meeting and discussion) was much less positive. 
The task turned out to be (1) much too difficult for novice users, and (2) VICToRy proved not to be of much help as it does not provide any information about why a rule is \emph{not} applicable in a specific situation (even though the expectation is that it should be).
While the possibility of selecting protocol entries and inspecting the resulting changes on the models was appreciated, many students stated that the opposite direction, i.e. selecting model elements in the visualisation and highlighting the ``responsible'' rule applications, would be indeed helpful as well and is currently missing.
Furthermore, there was a clear need for the introduction of breakpoints, i.e. the transformation should start in the background mode and stop at a certain point defined by the user, e.g., when a specific rule is applicable for the first time. 
This enables users to skip irrelevant parts of the transformation that are already clear to them and set the focus on debugging problematic steps.

Revisiting our research questions, our initial exploratory experiment at least indicates that (1) VICToRy appears to help obtain an overview  of a non-trivial \ac{TGG} the user is not familiar with, but (2) leaves room for improvement regarding the detection of errors in either models or rules.
An extension towards \textit{why-not-debugging} facilities~\cite{Anjorin2019} seems to be necessary to properly address bug finding tasks.
Our experiment is at best a pre-study for a more formal, controlled experiment and quantitative evaluation with multiple use cases, a larger group of test persons, and objective measures.
While we cannot generalise our results, our goal was not to provide hard empirical evidence for the effectiveness of VICToRy but rather to explore the design space in a realistic setting and brainstorm together with novice users for promising features to guide future extensions of VICToRy.
\section{Related Work}
\label{sec:related}

Several approaches to debugging in \ac{MDE} have been proposed, including fundamental concepts, debugging \ac{DSL} code, and debuggers for non-deterministic approaches.
Mierlo et al. describe a stepping semantics for debugging in \ac{MDE} with four levels of different granularity~\cite{Mierlo2018}.
The proposed approach is, however, conceptual and does not provide an implementation to the best of our knowledge.
A debugger for Petri nets is based on Modelverse and supports basic functionality including breakpoints known from \ac{GPL} debuggers~\cite{Mierlo2017b}.
The prototype is planned to be extended to support model transformations as well.
A wide range of facilities for \ac{DSL} debugging is presented in previous work.
Omniscient debugging - in contrast to stepwise execution - provides the user with enhanced navigation and exploration features such as reverting execution steps at runtime, impacting performance and scalability.
Therefore, approaches are often tailored to rather small instances~\cite{Corley2017} or specific use cases, such as xDMSLs (a subset of \acp{DSL})~\cite{Bousse2015}.
Lindeman et al. propose a declaratively defined debugger for \acp{DSL}~\cite{Lindeman2011}.
The approach was integrated into the Spoofax language workbench and evaluated by case studies involving the textual \acp{DSL} StrategoTL and WebDSL.
However, several limitations are mentioned for debugging modelling languages and model transformations.
Laurent et al. extended the \ac{fUML} by debugging facilities~\cite{Laurent2013}.
While the approach is a tool-independent add-on, it considers only the execution of models complying to the \ac{fUML} standard. 

For debugging rule-based systems, Tichy et al. sketch how to execute debugging steps for graph transformation, taking the tool Henshin as an example~\cite{Tichy2017}.
In contrast to our approach, the debugging of rule applications is much more detailed and takes the matching process into account as well, whereas an implementation is not described.
Similarly, Jukss et al. use graph transformations as an underlying formalism for a debugger integrated into AToMPM~\cite{Jukss2017}.
The approach focusses on a fine-grained inspection of the rule application process, whereas the user is not enabled to choose between multiple  possible rule applications.
For algebraic graph transformation, the tool AGG~\cite{Runge2011} provides a mode for stepwise execution of graph transformations.
Rule and match can be chosen by the user in each step, while it is neither clear which rules are applicable in the current state, nor a protocol of previous rule applications is provided.

Furthermore, a substantial number of tools have emerged that implement bidirectional model transformations based on \acp{TGG}.
While in Fujaba~\cite{Giese2006} and the TGG Interpreter~\cite{Greenyer2010}, operational rules are directly interpreted, MoTE~\cite{Hildebrandt2011,Giese2014}, eMoflon~\cite{Leblebici2014,Weidmann2019}, HenshinTGG~\cite{Ermel2012} and EMorF~\cite{Klassen2012} compile rules into source code of a \ac{GPL} to be executed at runtime.
A concept for debugging \acp{TGG} at different levels was introduced to the TGG Interpreter by Rieke~\cite{Rieke2015}.
The debugging facilities are, however, tightly interwoven with the specific tool and several open challenges for practical use are mentioned.
For MoTE, a monitor is implemented which allows to stepwise execute model transformations~\cite{Giese2014}.
However, the user cannot influence the execution order, which is determined by order of correspondence nodes in a processing queue and their respective types.
A debugging mode is implemented for EMorF as well, but both a detailed description and the tool itself are not currently available.
For all other \ac{TGG}-based tools, debugging functionality is missing to the best of our knowledge.

Besides these rule-based approaches, debugging plays an important role in other \ac{MDE}-related fields as well.
Proposed concepts include work on dynamic meta modelling~\cite{Bandener2010}, \acp{DEVS}~\cite{Mierlo2017c}, and story diagrams~\cite{Krasnogolowy2012}, which are each tailored to a specific tool and use case, though.
The tool TETRA Box is based on PaMoMo and involves white-box testing of transformation languages by symbolic execution of model transformations~\cite{Schoenboeck2013}, which is independent of the underlying transformation language but not yet tested with realistic examples.
SyVOLT localises errors in the input based on igraph and the T-Core framework~\cite{Oakes2018}, while the focus of debugging is set on detecting reasons for contract violations rather than on the transformation process. 
Ferdjoukh et al. localise faults in metamodel design based on static analyses and implemented their approach in TIWIZI and GRIMM~\cite{Ferdjoukh2018}, whereas model transformations are not taken into account.

\section{Conclusion and Future Work}
\label{sec:conclusion}

We presented the add-on component VICToRy for interactively visualising single steps of the model generation and transformation process.
Besides the inspection of possible rule applications in the current state, the user can inspect the prior transformation process using a transformation protocol.
The \ac{TGG}-based tool is fully integrated into the eMoflon tool suite but can be used along with other applications via a defined interface.
A concept for switching between background mode and debug mode via breakpoints is presented, whereas an implementation for breakpoints is left for future work.
Besides model generation and transformation, other consistency management operations such as model synchronisation and consistency checking are planned to be supported as part of future work.
To improve the tool's usability and effectiveness, support for easily defining different kinds of breakpoints should be implemented.
Furthermore - as an extension towards supporting why-not provenance - information about reasons for blocked rule applications should be presented to the user to support detecting logical faults in \ac{TGG} rules, or a mismatch with expectations in provided input models and tests.
 
\section*{Acknowledgments}

We would like to thank all members of the project group \textit{VICToRy} at Paderborn University for taking part in the tool development, namely Asher Ahsan, Philipp Giakoupian, Rifat Hussain, Jane Jose, Mahi Kittur, Israq Masrur, Hariprasath Ragupathy, Shubhangi
Salunkhe, Ayurshi Singh, Saman Soltani, Ankita Srivastava, Vipasyan Telaprolu, Mario Treiber, Surbhi Verma and Darya Zarkalam.

%

 \bibliographystyle{eptcs}
 \bibliography{references}

%
%
%
%
\end{document}